\documentstyle[12pt]{article}

\def\beq{\begin{equation}}
\def\eeq{\end{equation}}
\def\bea{\begin{eqnarray}}
\def\eea{\end{eqnarray}}

\def\ba{\begin{array}}
\def\ea{\end{array}}

\def\,{\"{U}}
\def\6{\.{I}}

\begin{document}

\title{Exact solution of Effective mass Schr\"{o}dinger
Equation for the Hulthen potential
 }
\author{\vspace{1cm}
{Ramazan Sever $^1$, Cevdet Tezcan $^2$,  \"{O}zlem Ye\c{s}ilta\c{s}
$^3$, Mahmut Bucurgat $^1$ }
         \\
{\small \sl  $^1$ Middle East Technical University,Department of
Physics, 06531 Ankara, Turkey}
\\{\small \sl $^2$ Faculty of Engineering, Ba\c{s}kent University,
Ba\~{g}l{\i}ca Campus, Ankara, Turkey }
\\ {\small \sl $^3$ Gazi University, Faculty of Arts and Sciences, Department of Physics,
06500, Ankara,Turkey}}
\date{\today}
\maketitle
\begin{abstract}
A general form of the effective mass Schr\"{o}dinger equation is
solved exactly for Hulthen potential. Nikiforov-Uvarov method is
used to obtain energy eigenvalues and the corresponding wave
functions. A free parameter is used in the transformation of the
wave function.
\\
{\small \sl \noindent PACS numbers: 03.65.-w; 03.65.Ge; 12.39.Fd \\[0.2cm]
}
\end{abstract}
\baselineskip 0.9cm
\newpage
\section{Introduction}
Quantum mechanical systems with  position dependent effective mass
(PDM) have been studied in different branches of physics by many
authors [1,2,3,4,5,6,7]. Several authors have obtained the exact
solutions of Schr\"{o}dinger equation with position dependent mass
[8-18]. Moreover, the Morse potential [19], one dimensional
Coulomb-like potential [20], hard core potential [21], harmonic
oscillator potential [22] are known as some real physical potentials
that have been investigated within PDM framework.

In recent years many authors have been used Nikiforov-Uvarov (NU)
approach for solving the Schr\"{o}dinger equation (SE)
[23,24,25,26,27,28,29].

In this work, the general form of PDEM Schr\"{o}dinger equation is
obtained by using a more general transformation of the wave
function as $\varphi=m^{\eta}(x)\psi(x)$. NU approach is adapted
to this general equation. Using an appropriate mass function, it
is solved for Hulthen potential within this generalization. Energy
eigenvalues and the corresponding wave functions are obtained. The
contents of the paper is as follows: in section II, we introduce
PDM approach and Nikiforov-Uvarov method. The next section
involves solutions of the general PDM equation. Results are
discussed in section IV.
\section{Method}
\noindent We write the one-dimensional effective mass Hamiltonian of
the SE as $\left[29\right]$
\begin{equation}
H_{eff}=-\frac{d}{dx}\left(\frac{1}{m(x)}\frac{d}{dx}\right)+V_{eff}(x)
\end{equation}
where $V_{eff}$ has the form
\begin{equation}
V_{eff}=V(x)+\frac{1}{2}(\beta+1)\frac{m^{''}}{m^{2}}-\left[\alpha(\alpha+\beta+1)+\beta+1\right]\frac{{m^{'}}^{2}}{m^{3}}
\end{equation}
with $\alpha$, $\beta$ are ambiguity parameters. Primes stand for
the derivatives with respect to $x$ and we have set
$\hbar=2m_{0}=1$. Thus the SE takes the form
\begin{equation}
\left(-\frac{1}{m}\frac{d^{2}}{dx^{2}}+\frac{m^{\prime}}{m}\frac{d}{dx}+V_{eff}-E\right)\varphi(x)=0
\end{equation}
We apply the following transformation
\begin{equation}
\varphi=m^{\eta}(x)\;\psi(x)
\end{equation}
Hence, the SE takes the form
\begin{equation}
\left\{-\frac{d^{2}}{dx^{2}}-(2\eta-1)\frac{m^{'}}{m}\frac{d}{dx}-
(\eta(\eta-2)+\alpha(\alpha+\beta+1)+\beta+1)\frac{m^{'2}}{m^{2}}+
(\frac{1}{2}(\beta+1)-\eta)\frac{m^{''}}{m}+m(V-E) \right\}\psi=0
\end{equation}
Hulthen potential is given by [26]
\begin{equation}
V(x) = -V_{0}\frac{e^{-\lambda x}}{1-qe^{-\lambda x}}
\end{equation}
We give the following parameters including mass relation:
\begin{eqnarray}
  A^{*} &=& \alpha(\alpha+\beta+1)+\beta+1 \\
    m(x) &=& (1-qe^{-\lambda x})^{-1}
\end{eqnarray}
\begin{eqnarray}
  \frac{m^{'}}{m} &=& -q\lambda \frac{e^{-\lambda x}}{1-qe^{-\lambda x}} \\
  \frac{m^{''}}{m} &=& q\lambda^{2}e^{-\lambda x}\frac{1+qe^{-\lambda x}}{(1-qe^{-\lambda x})^{2}}
\end{eqnarray}
We introduce a variable changing in Eq.(5)given as
\begin{equation}
s=\frac{1}{1-qe^{-\lambda x}}
\end{equation}
and if we use Eqs.(6),(7),(8),(9),(10) and (11) in Eq.(5), it
becomes,
\begin{equation}
\left(\frac{d^{2}}{ds^{2}}+\frac{2\eta-(2\eta+1)s}{s(1-s)}\frac{d}{ds}+
\frac{1}{s^{2}(1-s)^{2}}(-\xi_{1}s^{2}+\xi_{2}s-\xi_{3})\right)\psi=0.
\end{equation}
Parameters defined in Eq.(12) have the following form:
\begin{eqnarray}
  -\xi_{1} &=& (\eta(\eta-2)+A^{*})-2(\frac{1}{2}(\beta+1)-\eta)+\frac{V_{0}}{q\lambda^{2}} \\
  \xi_{2} &=& -2(\eta(\eta-2)+A^{*})+3(\frac{1}{2}(\beta+1)-\eta)-\frac{V_{0}}{q\lambda^{2}}+\frac{E}{\lambda^{2}} \\
  -\xi_{3} &=& \eta(\eta-2)+A^{*}-(\frac{1}{2}(\beta+1)-\eta).
\end{eqnarray}
where  $V(s)=\frac{V_{0}}{q}(1-s)$. Now, we apply the NU method
starting from its standard form
\begin{equation}
\psi^{''}_{n}(s)+\frac{\tilde{\tau}(s)}{\sigma(s)}\psi^{'}(s)+
\frac{\tilde{\sigma}(s)}{\sigma^{2}(s)}\psi_{n}(s)=0.
\end{equation}
Comparing Eqs.(12) and (16), we obtain
\begin{equation}
\sigma=s,  \tilde{\tau}(s)=3-4\eta,
\tilde{\sigma}(s)=-\xi_{1}s^{2}-\xi_{2}s+\xi_{3}
\end{equation}
In the NU method, the function $\pi$ and the parameter $\lambda$ are
defined as [23]
\begin{equation}
\pi(s)=\frac{\sigma^{'}-\tau(s)}{2}\pm
\sqrt{\left(\frac{\sigma^{'}-\tau(s)}{2}\right)^{2}-\tilde{\sigma}(s)+k\sigma(s)}
\end{equation}
and
\begin{equation}
\lambda=k+\pi^{'}
\end{equation}
To find a physical solution, the expression in  the square root
must be square of a polynomial. Then, a new eigenvalue equation
for the SE becomes
\begin{equation}
\lambda=\lambda_{n}=-n\tau^{'}-\frac{n(n-1)}{2}\sigma^{''}(s),
(n=0,1,2,...)
\end{equation}%
where
\begin{equation}
\tau(s)=\tilde{\tau}(s)+2\pi(s)
\end{equation}%
and it should have a negative derivative [23]. A family of
particular solutions for a given $\lambda$ has hypergeometric type
of degree. Thus, $\lambda=0$ will corresponds to energy eigenvalue
of the ground state, i.e. $n=0$. The wave function is obtained as a
multiple of two independent parts:
\begin{equation}
\psi(s)=\phi(s)y(s)
\end{equation}%
where $y(s)$ is the hypergeometric type function written with a
weight function $\rho$ as
\begin{equation}
y_{n}(s)=\frac{B_{n}}{\rho(s)}\frac{d^{n}}{ds}[\sigma^{n}(s)\rho(s)]
\end{equation}%
where $\rho(s)$ must satisfy the condition [23]
\begin{equation}
(\sigma \rho)^{'}=\tau \rho
\end{equation}%
The other part is defined as a logarithmic derivative
\begin{equation}
\frac{\phi^{'}(s)}{\phi(s)}=\frac{\pi(s)}{\sigma(s)}
\end{equation}%
\section{Solutions}
If we take Eq.(12) into account, comparing with Eq.(16), it is
observed that $\tilde{\tau}=2\eta-(2\eta+1)s$, $\sigma=s(1-s)$,
$\tilde{\sigma}=-\xi_{1}s^{2}+\xi_{2}s-\xi_{3}$. Using
$z=\frac{1}{2}(1-2\eta)$, one obtains
\begin{equation}
\pi=z(1-s)\pm \left\{%
\begin{array}{ll}
(\sqrt{\xi_{1}-\xi_{2}+\xi_{3}}-\sqrt{\xi_{3}+z^{2}})s+\sqrt{\xi_{3}+z^{2}},
& \hbox{$k_{1}=\xi_{2}-2\xi_{3}+2\zeta$;} \\
(\sqrt{\xi_{1}-\xi_{2}+\xi_{3}}
+\sqrt{\xi_{3}+z^{2}})s-\sqrt{\xi_{3}+z^{2}},
& \hbox{$k_{2}=\xi_{2}-2\xi_{3}-2\zeta$.} \\
\end{array}%
\right.
\end{equation}%
where
$\zeta=\sqrt{{\xi_{3}(\xi_{1}-\xi_{2}+\xi_{3}+z^{2})}-z^{2}(\xi_{2}-\xi_{1})}$.
Now we can introduce $\tau(s)$ as given below,
\begin{equation}
\tau(s)=\left\{%
\begin{array}{ll}
    1-2s+2((\sqrt{\xi_{1}-\xi_{2}+\xi_{3}}-\sqrt{\xi_{3}+z^{2}})s+\sqrt{\xi_{3}+z^{2}}) \\
    1-2s-2((\sqrt{\xi_{1}-\xi_{2}+\xi_{3}}-\sqrt{\xi_{3}+z^{2}})s+\sqrt{\xi_{3}+z^{2}}) \\
    1-2s+2((\sqrt{\xi_{1}-\xi_{2}+\xi_{3}}
+\sqrt{\xi_{3}+z^{2}})s-\sqrt{\xi_{3}+z^{2}}) \\
    1-2s-2((\sqrt{\xi_{1}-\xi_{2}+\xi_{3}}
+\sqrt{\xi_{3}+z^{2}})s-\sqrt{\xi_{3}+z^{2}}) \\
\end{array}%
\right.
\end{equation}
Derivative of $\tau(s)$ is obtained as
\begin{equation}
\tau^{'}=\left\{%
\begin{array}{ll}
    -2+2(\sqrt{\xi_{1}-\xi_{2}+\xi_{3}}-\sqrt{\xi_{3}+z^{2}}) \\
    -2-2(\sqrt{\xi_{1}-\xi_{2}+\xi_{3}}-\sqrt{\xi_{3}+z^{2}})\\
    -2+2(\sqrt{\xi_{1}-\xi_{2}+\xi_{3}}+\sqrt{\xi_{3}+z^{2}}) \\
   -2-2(\sqrt{\xi_{1}-\xi_{2}+\xi_{3}}+\sqrt{\xi_{3}+z^{2}}) \\
\end{array}%
\right.
\end{equation}
Here, first derivative of $\tau$ should be $\tau^{'}< 0$ in order to
obtain physical solutions. Thus we choose $k$ and our functions
which help us to derive the energy eigenvalues and eigenfunctions:
\begin{eqnarray}
  k &=& \xi_{2}-2\xi_{3}-2\sqrt{{\xi_{3}(\xi_{1}-\xi_{2}+\xi_{3}+z^{2})}-z^{2}(\xi_{2}-\xi_{1})} \\
  \tau &=& 1-2s-2[(\sqrt{\xi_{1}-\xi_{2}+\xi_{3}}+\sqrt{\xi_{3}+z^{2}})s-\sqrt{\xi_{3}+z^{2}}] \\
  \pi &=& z(1-s)-[(\sqrt{\xi_{1}-\xi_{2}+\xi_{3}}+\sqrt{\xi_{3}+z^{2}})s-\sqrt{\xi_{3}+z^{2}}] \\
  \tau^{'} &=&
  -2-2(\sqrt{\xi_{1}-\xi_{2}+\xi_{3}}+\sqrt{\xi_{3}+z^{2}}).
\end{eqnarray}
Using Eq.(19), the relation given below
\begin{equation}
\lambda=z^{2}-z+\xi_{1}-(\sqrt{\xi_{1}-\xi_{2}+\xi_{3}}+\sqrt{\xi_{3}+z^{2}})^{2}-
(\sqrt{\xi_{1}-\xi_{2}+\xi_{3}}+\sqrt{\xi_{3}+z^{2}})
\end{equation}%
is obtained. With the aid of Eq.(20), this equality can be
written:
\begin{equation}
\lambda=\lambda_{n}=-n(-2-2(\sqrt{\xi_{1}-\xi_{2}+\xi_{3}}+\sqrt{\xi_{3}+z^{2}}))+n(n-1)
\end{equation}%
Substituting
$\Lambda=\sqrt{\xi_{1}-\xi_{2}+\xi_{3}}+\sqrt{\xi_{3}+z^{2}}$,
$\Lambda$ can be written
\begin{equation}
\Lambda=\frac{1}{2}\left(-(2n+1)\pm\sqrt{1+4\gamma}\right)
\end{equation}%
where $\gamma=\xi_{1}+z(z-1)$. Now let us discuss two cases here
depending on signs of $\Lambda$.
\\
Case 1:
\begin{equation}
\sqrt{\xi_{1}-\xi_{2}+\xi_{3}}+\sqrt{\xi_{3}+z^{2}}=\frac{1}{2}\left(-(2n+1)+\sqrt{1+4\gamma}\right)
\end{equation}%
then, $\xi_{3}$ is obtained as
\begin{equation}
\xi_{3}=\left(\frac{\xi_{2}-\xi_{1}+z^{2}}{2n+1-\sqrt{1+4\gamma}}+
\frac{1}{4}\left(2n+1-\sqrt{1+4\gamma}\right)\right)^{2}
\end{equation}%
\\
Using the definitions of $\xi_{1}, \xi_{2}$ and $\xi_{3}$, $E_{n}$
is given by
\begin{equation}
E_{n}=-\frac{\lambda^{2}}{4}\left(2n+1-\sqrt{1+4\gamma}-
2\sqrt{-\eta(\eta-1)-A^{*}+\frac{\beta+1}{2}}\right)^{2}-\lambda^{2}(\eta-\frac{1}{2})^{2}
\end{equation}%
Case 2:
\begin{equation}
\sqrt{\xi_{1}-\xi_{2}+\xi_{3}}+\sqrt{\xi_{3}+z^{2}}=\frac{1}{2}\left(-(2n+1)-\sqrt{1+4\gamma}\right)
\end{equation}%
then, $\xi_{3}$ reads
\begin{equation}
\xi_{3}=\left(\frac{\xi_{2}-\xi_{1}+z^{2}}{2n+1+\sqrt{1+4\gamma}}-
\frac{1}{4}\left(2n+1+\sqrt{1+4\gamma}\right)\right)^{2}
\end{equation}%
Energy eigenvalues can be written:
\begin{equation}
E_{n}=\frac{\lambda^{2}}{4}\left(2n+1+\sqrt{1+4\gamma}+
2\sqrt{-\eta(\eta-1)-A^{*}+\frac{\beta+1}{2}}\right)^{2}-\lambda^{2}(\eta-\frac{1}{2})^{2}
\end{equation}%
Using Eqs(24) and (25), $\phi$ and $ \rho$ are obtained as
\begin{equation}
\phi=s^{z+\sqrt{\xi_{3}+z^{2}}}(1-s)^{\sqrt{\xi_{1}-\xi_{2}+\xi_{3}}}
\end{equation}%
and
\begin{equation}
\rho(s)=s^{2\sqrt{\xi_{3}+z^{2}}}(1-s)^{2\sqrt{\xi_{1}-\xi_{2}+\xi_{3}}}
\end{equation}%
Solution of $y$ can be obtained from Eq.(23):
\begin{equation}
y_{n}(s)=P^{(2\sqrt{\xi_{3}+z^{2}},2\sqrt{\xi_{1}-\xi_{2}+\xi_{3}})}_{n}(1-2s).
\end{equation}%
Hence, the wave function has the following form:
\begin{equation}
\psi_{n}=s^{z+\sqrt{\xi_{3}+z^{2}}}(1-s)^{\sqrt{\xi_{1}-\xi_{2}+
\xi_{3}}}P^{(2\sqrt{\xi_{3}+z^{2}},2\sqrt{\xi_{1}-\xi_{2}+\xi_{3}})}_{n}(1-2s)
\end{equation}%
If $z+\sqrt{\xi_{3}+z^{2}} < 0$ and $\sqrt{\xi_{1}-\xi_{2}+
\xi_{3}} > 0$, it is required that  $|z+\sqrt{\xi_{3}+z^{2}}|\geq
\sqrt{\xi_{1}-\xi_{2}+ \xi_{3}}$ and if $\sqrt{\xi_{1}-\xi_{2}+
\xi_{3}} < 0$, $z+\sqrt{\xi_{3}+z^{2}} > 0$,
$|\sqrt{\xi_{1}-\xi_{2}+ \xi_{3}}| \geq z+\sqrt{\xi_{3}+z^{2}}$
for physical solutions.
\section{Conclusions}
NU method adapted solutions are obtained for Hulthen potential
within PDEM Schr\"{o}dinger equation. We have proposed a
transformation of the wavefunction in a general form that leads to
solutions of well-known eigenvalues and eigenfunctions of Hulthen
potential. Furthermore, energy relations of the mass independent
equation are obtained for two cases.

\section{Acknowledgements}
This research was partially supported by the Scientific and
Technological Research Council of Turkey.


\newpage

\begin{thebibliography}{99}
\bibitem{ref1} G Bastard, Wave Mechanics Applied to Semiconductor Heterostructures (Les Ulis:
Editions de Physique), 1998.
\bibitem{ref2} L I Serra, E Lipparini, Europhys. Lett. 40 1997
667.
\bibitem{ref3} M Barranco, M Pi, S M Gatica, E S  Hernandez, J Navarro, Phys. Rev. B 56
1997 8997.
\bibitem{ref4} F de Saavedra Arias et al, Phys. Rev. B 50 1994
4248.
\bibitem{5} Von Roos O 1983 Phys. Rev. B 27 7547.
\bibitem{6} L Dekar, L Chetouani, T F  Hammann, J. Math. Phys. 39 (1998)
2551.
\bibitem{7} A R  Plastino, A Rigo, M  Casas, F Garcias, A  Plastino, Phys. Rev. A 60 (1999) 4318
\bibitem{8} A Ganguly, L M Nieto, J. Phys. A, 40 (2007) 7265;
Chun-Sheng Jia, Jian-Yi Liu, Ping-Quan Wang, Chao-Shan Che, Phys.
Lett. A,  369 (2007) 274.
\bibitem{9} B Bagchi, J. Phys. A: Math. Theor. 40 (2007) F1041.
\bibitem{10} T Tanaka, J. Phys. A: Math. Gen. 39 2006 219-234.
\bibitem{11} B Roy, P Roy, J. Phys. A 35 (2002) 3961.
\bibitem{12} B Roy, Europhys. Lett. 72 2005 1-6;  R Ko\c{c}, M Koca, J. Phys. A: Math. Gen. 36 (2003)
8105–8112.
\bibitem{13} C Quesne, SIGMA 3 (2007) 067.
\bibitem{14} G Chen, Zi-dong Chen, Phys. Lett. A, 331(5) 2004
312-315.
\bibitem{15} R De, R Dutt and U Sukhatme, J. Phys. A, 25(13) 1992
L843-L850.
\bibitem{16} A D Alhaidari, Phys. Rev. A 66 2002 042116.
\bibitem{17} Shang-Wu Qian et al, New J. of Physics, 4 2002 13.1-13.6;
Chen Gang, Chinese Phys. 14 2005 460-462.
\bibitem{18} B Bagchi, P Gorain, C
Quesne, R Roychoudhury, Mod. Phys. Lett. A 19 2004 2765.
\bibitem{19} J Yu, S H  Dong, G H  Sun, Phys. Lett. A, 322 (2004)
290.
\bibitem{20} J Yu, S H  Dong,  Phys. Lett. A, 325 (2004) 194.
\bibitem{21} S H  Dong, M  Lozada-Cassou, Phys. Lett. A, 337 (2005)
313.
\bibitem{22} L  Jiang, L Z  Yi, C S  Jia, Phys. Lett. A,
345 (2005) 279.
\bibitem{23} A F Nikiforov and V B Uvarov, Special Functions of Mathematical Physics
(Birkhauser, Bassel, 1988).
\bibitem{24} H E\~{g}rifes, D Demirhan, F B\"{u}y\"{u}kkili\c{c}, Phys. Lett. A, 275(4) 2000
229-237; H E\~{g}rifes, D Demirhan, F B\"{u}y\"{u}kkili\c{c},
Phys. Scr. 59 No 2 (1999) 90-94.
\bibitem{25} \"{O} Ye\c{s}ilta\c{s} M. \c{S}im\c{s}ek, R. Sever,
C. Tezcan, Phys. Scr. 67  (2003) 472-475.
\bibitem{26} H E\~{g}rifes, R Sever, Phys. Lett. A, 344(2-4) 2005
117-126.
\bibitem{27}  H E\~{g}rifes, R Sever, Int. J. of Theo. Phys., 46(4)
2007935.
\bibitem{28}  S M Ikhdair, R Sever, arXiv:quant-ph/0605045v1
\bibitem{29} B. Bagchi, P.S. Gorain, C. Quesne, Mod. Phys. Lett. A 21 (2006)
2703-2708.


\end{thebibliography}
\end{document}